# State-of-the-Art Survey on In-Vehicle Network Communication "CAN-Bus" Security and Vulnerabilities

[1]**Omid Avatefipour**, [2]**Hafiz Malik**

[1] University of Michigan – Dearborn
Department of Electrical and Computer Engineering
Dearborn, Michigan, United States
*oavatefi@umich.edu*

[2] Associate Professor, University of Michigan – Dearborn
Department of Electrical and Computer Engineering
Dearborn, Michigan, United States
*hafiz@umich.edu*

**Abstract -** Nowadays with the help of advanced technology, modern vehicles are not only made up of mechanical devices but also consist of highly complex electronic devices and connections to the outside world. There are around 70 Electronic Control Units (ECUs) in modern vehicle which are communicating with each other over the standard communication protocol known as Controller Area Network (CAN-Bus) that provides the communication rate up to 1Mbps. There are different types of in-vehicle network protocol and bus system namely Controlled Area Network (CAN), Local Interconnected Network (LIN), Media Oriented System Transport (MOST), and FlexRay. Even though CAN-Bus is considered as de-facto standard for in-vehicle network communication, it inherently lacks the fundamental security features by design like message authentication. This security limitation has paved the way for adversaries to penetrate into the vehicle network and do malicious activities which can pose a dangerous situation for both driver and passengers. In particular, nowadays vehicular networks are not only closed systems, but also they are open to different external interfaces namely Bluetooth, GPS, to the outside world. Therefore, it creates new opportunities for attackers to remotely take full control of the vehicle. The objective of this research is to survey the current limitations of CAN-Bus protocol in terms of secure communication and different solutions that researchers in the society of automotive have provided to overcome the CAN-Bus limitation on different layers.

## 1. Introduction

The Controller Area Network (CAN-Bus) protocol was introduced in 1983 by Robert BOSCH GmbH and has been widely applied in the automotive communication and even in domestic appliances, medical devices, and entertainment domains. [1]. Compared to the TCP/IP protocol in which the origin and destination addresses are defined in each packet, CAN-Bus messages does not have origin and destination address and instead it utilizes the broadcasting communication technique in such a way that each node in network can send and receive packets to/from bus.

Since there is no destination address in CAN-Bus, each node can publish and receive particular messages based on the pre-defined node (here ECU) configuration. This communication technique increases the network elasticity [2] which means that if new ECU is supposed to add to the current network, it will be configured easily and does not require any changes to the network infrastructure and other nodes as well. CAN-Bus is event trigger protocol which means a message is generated in reply to the generation of event or request in the network. CAN-Bus is considered as multi-master protocol that defines if the communication bus is free, any node can publish/receive message on the bus and the latency time is guaranteed as well. Vehicular network has introduced a variety of merits such as reducing harness in large extent, establishing data sharing, remarkably improving the intelligent control level of vehicle e.g. Advanced Driving Assistant Systems (ADAS), improving capabilities of failure diagnosis and repair and so on.

To meet the real-time systems deadline requirements, each message has been assigned an identifier frame which is utilized to define the message priority [3]. The lower number of message identification value, the higher priority it has to gain the bus. This prioritization feature has also solved the bus access conflict in such a way that if two nodes want to send data simultaneously, each ECU which has a lower ID value will publish the message firstly. (Due to the higher priority). This technique is also known as message arbitration. [2]. Generally, CAN regulates arbitration in a predictable and efficient manner. Figure 1 depicts a situation that three nodes (First node: 11001011111 in binary, second node: 110011111111 in binary, and third node 110010110010 in binary) try to transmit message simultaneously. In order to prevent bus collision, a given node with the lowest ID (in this case third node) will transmit the information because it has a lowest value and highest priority than the other two nodes. Figure 1 depicts the message arbitration of this scenario.

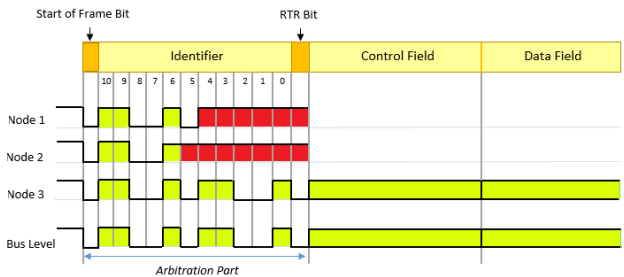

Fig. 1. Arbitration condition in CAN-Bus protocol.

The rest of this paper is organized as follows: Section 2 presents an overview of CAN-bus protocol. Section 3 discusses the CAN-Bus protocol vulnerabilities. Section 4 provides a state-of-the-art survey in the area of CAN-Bus security and authentication techniques and finally the paper is concluded in Section 5.

## 2. CAN-Bus Protocol – an overview

Before discussing the security vulnerabilities of CAN-Bus protocol, in this section an overview of CAB-Bus protocol is provided. Generally, there are two formats of CAN-Bus: standard format which has 11-bit for identifier and extended-format which includes 29-bit identifier frame. Data Frame, Remote Frame, overload frame, and error frame are four major frame types in controlled area network (CAN-Bus). Data Frame is used to carry the data from a transmitter to a receiver, which consists of the following bit fields: start of frame, (one dominant bit), arbitration field which consists of 12 bits, control field which has 6 bit, and data field (in range of 0 to 64 bytes), CRC field (16-bit), ACK field (2-bit), and End of Frame (7-bit). The complete illustration of data frame is shown in Figure 2. Arbitration field defines the priority of each message and also there is a single bit in this field to define whether this is a data frame or remote frame. [4] Remote frame is used to enable the receiver to request another data from transmitter. The data frame can be in the length of zero (remote frame) to eight bytes and control field specifies the length of the data frame. CRC frame: CRC frame consists of 16 bits totally; 15 bits are used for Cyclic Redundant Checksum algorithm for error detection and one recessive bit as delimiter. ACK field: Receiver node re-computes the CRC and if it matches, it reports this to the transmitter that the valid message has been received correctly. It is done by overwriting the recessive bit (logic 1) in ACK slot with the dominant bit (logic 0) [4].

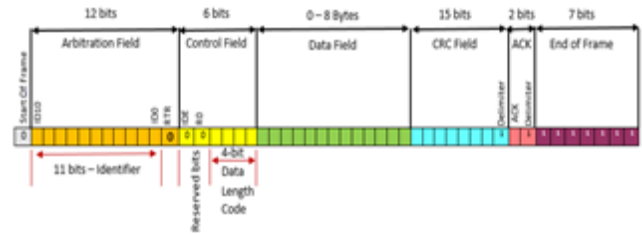

Fig. 2. CAN-Bus data frame

Bit stuffing technique is used in CAN-Bus which indicates that if there are six consecutive identical bits transmitted in the bus, it is considered as an error because bit stuffing law is violated [5]. Bit stuffing can be applied in different frames in CAN-Bus e.g. arbitration field, control field, and CRC field which means a complementary bit will be added to the frame when the transmitter finds that there are five identical bits consecutively. Therefore, six consecutive identical bits during the transmission is considered as bit-stuffing violation and error frame will be transmitted by each node which detects this situation. In Figure 3 the CAN frame and how bit stuffing is applied to the frame is shown.

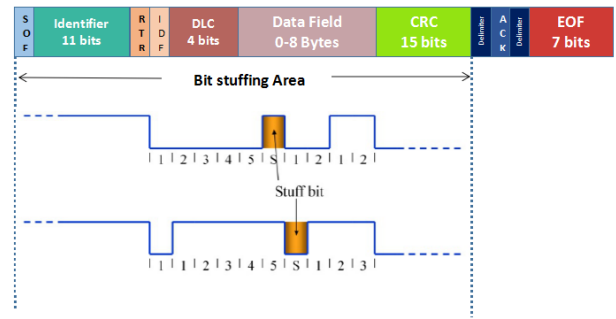

Figure 3. Bit stuffing technique in CAN-Bus.

CAN-Bus consists of three major layers namely physical layer, transfer layer, and object layer. Physical layer includes the actual bit transfer between different nodes and the electrical properties of transmission and also the medium which is used for communication [6].

CAN-Bus communication channel consists of twisted pair wires knowns as CAN-High and CAN-Low. There are two different state for bit transmission e.g. recessive logical "1" and dominant logical "0". After the CAN packet is passed from the CAN interface, the CAN transceiver (transmitter/receiver) converts the packet into a differential signal for transmission over the twisted pair wire. Most of the automotive communication protocol utilize the physical differential signaling specified in ISO 11898-2 – High speed CAN up to 1Mbps [7]. When a recessive bit (logical 1) is transmitted, both CAN-High and CAN-Low wires carry 2.5 V which means the voltage difference becomes zero On the other hand, during the transmission of dominant (logical 0) bit, CAN-High increases its voltage by 1 V, resulting in 3.5 V on CAN-High and CAN-Low voltage decreases its voltage by 1, resulting in 1.5 V on CAN-Low which means the voltage difference between CAN-High and CAN-Low becomes 2 volts. Figure 4 illustrates CAN-Bus voltage level. In this way the average voltage on the wire is always 2.5 V, making CAN-Bus very resilient against electric and magnetic interference.

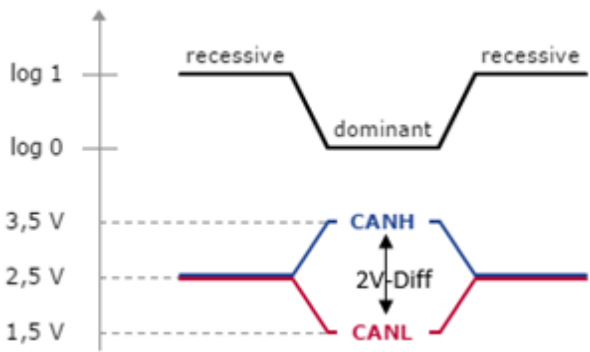

Figure 4. CAN-Bus differential signal illustration.

Connecting the vehicular networks to different environments, both internal networks and wireless, creates fantastic services for the automotive industry in terms of efficiency, cost and safety e.g. Vehicle to Vehicle (V2V), and Vehicle to Infrastructure (V2I) communication, Firmware-Update-Over-The-Air (FOTA) and remote diagnostics that enables embedded software components to be re-programed remotely and provides advantages for drivers in a way that they do not need to bring the vehicle to dealer for diagnostic services. [8]. However, these features can introduce new challenges because both internal and external communication needs to be secured properly otherwise attackers can take full control of the vehicle and endanger the passengers' life consequently by misusing these features.

## 3. CAN-Bus Protocol Vulnerabilities

The CAN-bus protocol was designed to be lightweight, robust, and fast as it should be capable of having satisfactory performance in real-time environment and meet the time constraints [9]. However, CAN-Bus contains several vulnerabilities which are included in its design and has paved the way for adversaries to have access to the network and inject malicious message for different purposes. From security solution standpoint, a secure communication should meet these five criteria by protocol or system security designer [10]:

- *Data Integrity:* information which is received by the receiver should be exactly the same as sender has sent in channel without any alternation.

- *Authentication:* all parties (ECUs in CAN-Bus) should be detected that they are authenticated.

- *Confidentiality:* the communication between authorized parties should be protected against unauthorized ones.

- *Nonrepudiation:* the security solution should prove that the parties in the communication cannot deny the authenticity of the message that was organized.

- *Availability:* the security solution should ensure that the system availabilities throughout different circumstances are guaranteed.

One of the inherent limitation of CAN-Bus, which makes the nodes in network to be compromised, is the lack of message authentication within each CAN message. As its name implies, CAN-Bus is a network of different controllers with different functionalities. For instance, Engine Control Unit is sending the RPM data continuously to the bus and it becomes available for all the nodes in CAN-Bus, irrespective of whether nodes in the bus have requested that message or not. The other nodes constantly listen to the bus for their specific message which can be recognized by the message identifier. The CAN-Bus architecture works fine in the normal circumstances. However, it does not provide security facilities by design to prevent unauthorized node from joining the communication and broadcast malicious messages to other nodes. These inherent vulnerabilities give the attacker a potential surface to send spoofed message after understanding the legitimate format of CAN-Bus, and each

ECU can impersonate the other ECUs for replay attack which could create harmful consequences for vehicle occupant. Attackers passively listen to the bus to record different legitimate messages content for different functionalities and then he/she can inject their own messages to manipulate the vehicle functionalities [11].

Another vulnerability of the CAN-Bus protocol is the unencrypted traffic during the communication. Encryption techniques never apply during the phase of protocol design since they can make overhead for real-time communication and this would be in contrast with the nature of the protocol (lightweight and fast). This problem makes surface straightforward for adversaries to sniff the traffic by simply buying a low-price hardware which can be connected to the CAN-Bus and passively sniff data and obviously without some forms of encryptions, message authenticity and integrity would not guarantee and then be able to perform malicious activities. Therefore, it is required to add some security level or plug-in to the current protocol to avoid these incidents [11].

Misuse of protocol is another reason that hacker can take advantage of it. For instance, as mentioned in the earlier part, CAN-Bus uses message arbitration to win the bus for data broadcasting when more than one node tries to send the data. A Denial-of-Service (DoS) attack can be launched by using the message arbitration technique in a way that adversary sends a malicious message with the highest priority (lowest ID) continuously. Therefore, the data-bus will be occupied all the time by the compromised node and could resulted in system failure [12]. Nowadays, by emerging the machine learning and intelligent algorithms, several methods are proposed in various engineering application e.g. intelligent controller design for industrial robots [13-15], Intrusion Detection Systems (IDS) [16-18], adaptive optimization algorithm [19-21], etc. Machine learning algorithms have been widely used as a powerful mathematical tool to develop security solution in the area of vehicular networks [22-25].

## 4. Related Work
In this part, the state-of-the-art survey is carried out to discuss different approaches and solutions that researchers have proposed to make in-vehicle communication more secure. Researchers have worked in different CAN-Bus layers to introduce security solutions. Cho and Shin [26] proposed a clock skew based framework for ECU fingerprinting and use it for the development of Clock based Intrusion Detection System (IDS). The proposed clock based fingerprinting method [26] exploited clock characteristic which exists in all digital systems: "*tiny timing error known as clock skew*". The clock skew identification exploits uniqueness of the clock skew and clock offset which is used to identify a given ECU based on clock attributes of the sending ECU. The proposed method measures and leverages the periodic behavior of CAN-Bus messages to fingerprint each ECU in the network and then constructing a reference clock behavior of each ECU by using Recursive Least Square (RLS) algorithm. Based on the developed reference behavior, deviation from the baseline clock behavior would consider as abnormal behavior (ECU is compromised) with low rate of false positive error: 0.055%. Cho and Shin developed a prototype for the proposed IDS and demonstrated effectiveness of the proposed CIDS on three different vehicles e.g. Honda Accord, Toyota Camry, and a Dodge Ram.

Wang et al [27]. propose a practical security framework for vehicular systems (VeCure), which can fundamentally solve the message authentication issue of the CAN bus. They validate the proposed method by developing a proof-of-concept prototype using Fessscale automotive development board. In their method each node which sends a CAN packet needs to send the message authentication code packet (8 bytes) as well. They divided the ECUs into two categories namely Low-trust group and High-trust group. ECUs which have external interfaces e.g. OBD-II or telematics are put in the low-trust group. High-trust group share a secret symmetric key to authenticate each coming and outgoing messages in a way that an ECU from Low-trust group that does not know the key cannot send message to critical ECUs in high-trust group. Wang et al. used SHA-3 hash function but they improve the system throughput by pre-calculating of the heavy weighted cryptographic function. The proposed method creates 2000 additional clock cycle compared to the system without message authentication technique (equals to the 50 micro second by running on the 40 MHz processor). By offline pre-calculating the hash function their method is 20-fold faster computationally than the other methods which uses message authentication solutions. Figure 5 depicts the proposed method, CAN-Bus without message authentication, and classic SHA-3 hash function in terms of number of CPU clock cycles that they consume.

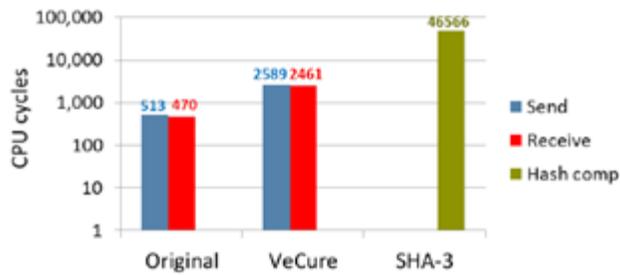

Figure 5. No. of CPU cycle of CAN-Bus without message authentication, VeCure, and classic SHA-3 hash function.

Koscher et al. [28] carried out a comprehensive experimental analysis of vehicle attack surfaces. They have analyzed different threat models and vulnerabilities with different range of vectors e.g. diagnostics mechanics sessions in which the adversary has a physical access to the bus via OBD-II port and by running a program on laptop to inject malware to the CAN-Bus. Infotainment systems in modern car have introduced several fascinating features e.g. connecting to the internet, cellphone, importing all the cellphone log to the infotainment screen like contact lists, etc. These features open a new surface for attackers to inject the malware in an audio file and by playing the modified audio file, the infotainment systems can be comprised and finally the attacker can steal the logged data which have saved at infotainment systems. Koscher et al. also examine both short rage wireless access e.g. Bluetooth, remote key less entry, RFID, and long range wireless e.g. GPS and satellite radio. They perform different attacks with the help of these surfaces. For instance, they manipulated the WMA audio file in a way that it is played perfectly on PC. However, in the background it sends CAN-Bus messages when the CD is played by the victim vehicle. The question which might come up to the mind is that why car manufactures do not consider these vulnerabilities during the CAN-Bus development? Koscher et al. discussed that vehicles had not been targeted for these types of attacks and on that time there were not as diverse surfaces of communication as we have nowadays. But vehicles nowadays as connected with several short-range and large-range wireless network and by introducing V2V & V2I communication this trend is continuously growing and consequently the opportunities for attackers would be more provided and in-vehicle network vulnerabilities will be increased as well.

Paar et al. [29] researchers from the Germany presented that the remote keyless entry which is becoming predominant feature for modern vehicles can be comprised and they can break the system based on the Keeloq RFID technology. This vulnerability can be applied to all remote keyless entry or other remote building access control systems which use Keeloq as cipher. They showed that the key less remote access can be compromised from a distance of 100 meters. Theoretically, the car generates random value which will be processed by the remote keyless module and by matching the correct calculation the car door will be open. Replay attacks are not allowed by the security protocol that even an adversary records all communication between two parties and try to impersonate one of the parties later on, the replay of the log file does not allow him to open the door. However, Paar et al. applied the side channel attack of these systems.

Hoppe et al. [30] performed four different tests on the control of window lift, warning light, airbag control systems and central gateway. They also classified and summarized their result in the CERT taxonomy for the security penetration and vulnerabilities of each part and analyze two selected counter-measures. They provide some short-term and long-term solution and believe the short-term solutions can adopt into the current vehicle electronic systems but for the long-term solution some major alternation in the protocol design is required. For instance, intrusion detection systems (IDS) and data analysis is introduced as short-term security solution. In the first scenario the electric window lift is targeted in the CANoe (simulation software by Vector CANTech company) in which the vehicular network is simulated and when a predefined condition is met (car speed goes beyond 200 km/h) by adding some lines of malicious codes, the electric window lift automatically is opened and will not close until the end of attack. This attack lies down on the "Read" and "Spoof" method to monitor the current traffic and when it reaches the specified condition, it spoofs the command for electric window lift and finally Denial of Service will be performed and does not allow driver to halt the attack when it is running. Hackers use the vulnerabilities of CAN-Bus since the messages are not authenticated during the communication and the malicious code is sent from the unauthorized ECU.

For the second scenario, Hoppe et al. target the warning lights (indicators). In the normal circumstances, when unauthorized opening of a door happens, the corresponding door sensor will send message to the ECU and some events will be triggers e.g. generating light and horn alarm for a couple of seconds. In this scenario when hacker opens vehicle door, the triggered "on" alarm will be set to "off" immediately which leads to turning off the light bulb and horn switched off and thief can steal the car or the items from the interior without any alarm. Again this vulnerability lies down on the CAN-Bus architecture communication (no message authentication) and this is "read" and "spoof" attack action and Denial of Service (DoS) as well. In the third scenario Hoppe et al. analyze the air bag control system. In this attack scenario, the air bag module will be removed from the system which leads to dire consequence during the car accident (air bog does not work in emergency cases). They believe that the intention of this attack can be monetary goals because after air bag deploys in the accident, its substitution could be costly. This attack scenario can be done by a compromised powertrain subnetwork ECU or by connecting a hardware to the OBD-II port. Additionally, they controlled the air bag controller indicator that does not indicate the air bag failure anymore. In table 1 the CERT classification of three aforementioned scenarios are summarized:

Table 1. CERT Classification of three attack scenarios

| Scenario | Attacker | Vulnerability | Action | Target | Result |
|---|---|---|---|---|---|
| Electric window system | Hackers By injecting the malicious code | CAN bus protocol no message authentication | Read/ spoof | Control Unit (e.g. right door) | Blocking of the window system (DoS) |
| Warning lights (indicators) | Thieves by injecting malicious code | CAN bus protocol no message authentication | Read/ spoof | Control Unit (ECU) | Blocking of the warning light system (DoS) |
| Air bag control system | Re-seller By injecting malicious code (OBD-II) port | CAN bus protocol no message authentication | Read/ Spoof Copy | Air bag ECU | Theft of resources (airbag function) |

One of the short-term countermeasure is developing the Intrusion Detection Systems (IDS). When a malicious activity or network pattern is detected by an intelligent detection system, it should create some alarm or warning to limit the consequences of the attack. e.g. stop the car at the next safe position. One capability that an IDS is detecting the message frequency. For instance, in scenario 1 & 2 the corresponding messages send in a constant frequency from a specified identifier. Attacker basically tries to send the exact identifier but with different content. Since removing the existing message is hard to achieve, therefore adversary will try to send the altered message with the same identifier within the significantly higher frequency. Hence, if the IDS can detect the high frequency of suspicious activity, it can create some warning alarm to the driver accordingly.

## 5. Conclusion

In this study, in-vehicle network communication protocol CAN-Bus and its corresponding vulnerabilities are introduced. Several researchers have performed to show its corresponding weaknesses in terms of penetrations to the network. Although some researchers proposed security solutions for the current protocol, most of the work in this area are carried out to introduce the current problems and their solutions are not comprehensive. Developing security solution in physical layer security would have more merits compared to transfer layer because one of the challenges for developing security mechanism in the transfer layer (applying message authentication code) is the limitation of computational power and memory of the microcontrollers which could be insufficient to develop a cryptographic algorithm for CAN-Bus in real-time environment. Vehicles are being revolutionized by integrating modern computing and communication technologies in order to improve both user experience and driving safety. As a result, vehicular systems that used to be closed systems are opening up various interfaces, such as Bluetooth, 3G/4G, GPS, etc., to the outside world. One of the root vulnerabilities of CAN-Bus is that there is no message authentication code for the communication which paves the way for adversaries to penetrate to the bus. It seems that by introducing the V2V and V2I communication, which increase external interfaces, the in-vehicle network communication protocol needs to be redesigned to boost the security aspect of the protocol or the more secured protocol e.g. Ethernet should be utilized more for in-vehicle network communication.

selected short-term countermeasures. *Computer Safety, Reliability, and Security*, 235-248.

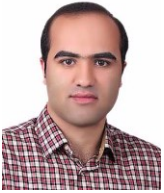

**Omid Avatefipour** is currently pursuing his Master's program in Computer Engineering at University of Michigan-Dearborn. His research interests include in-vehicle network communication protocol security, Embedded Systems, Data mining, Intelligent Control systems and Robotics. He has work experience at Vector CANTech company as Embedded Software Engineer and at Valeo North Amercia company as System Software Engineer in Advanced Engineering Research & Development department. He has also worked as researcher in Information System, Security, and Forensics (ISSF) laboratory at Department of Electrical and Computer Engineering (ECE), University of Michigan – Dearborn. Additionally, he was working as primary researcher in the laboratory of Control and Robotics at institute of Advanced Science and Technology, IRAN SSP Research & Development center.

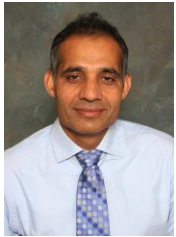

**Hafiz Malik** is Associate Processor in the Electrical and Computer Engineering (ECE) Department at University of Michigan – Dearborn. His research in cybersecurity, multimedia forensics, information security, wireless sensor networks, steganography/ steganalysis, pattern recognition, information fusion, and biometric security is funded by the National Academies, National Science Foundation and other agencies. He has published more than 70 papers in leading journals, conferences, and workshops. He is serving as Associate Editor for the IEEE Transactions on Information Forensics and Security since August 2014 and for the Springer Journal of Signal, Image, and Video Processing (SIVP) May 2013 – present. He is also on the Review Board Committee of IEEE Technical Committee on Multimedia Communications (MMTC). He organized Special Track on Doctoral Dissertation in Multimedia, in the 6th IEEE International Symposium on Multimedia (ISM) 2006. He is also organizing a special session on "Data Mining in Industrial Applications" within the IEEE Symposium Series on Computational Intelligence (IEEE SSCI) 2013. He is serving as vice chair of IEEE SEM, Chapter 16 since 2011. He is also serving on several technical program committees, including the IEEE AVSS, ICME ICIP, MINES, ISPA, CCNC, ICASSP, and ICC. He is a senior IEEE member.